\documentclass{appolb}
\usepackage[utf8]{inputenc}
\usepackage{cite}
\usepackage{paralist}
\usepackage{graphicx}

\newcommand{\dd}{\mathrm{d}}

\begin{document}

\title{Probing the QCD phase diagram with dileptons --- a study using coarse-grained transport dynamics
\thanks{Presented at CPOD 2016, Wroclaw, Poland, May 30 -- June 4, 2016}%
}
\author{Stephan Endres, Hendrik van Hees, and Marcus Bleicher
\address{Frankfurt Institute for Advanced Studies, Ruth-Moufang-Straße 1, \\ 60438 Frankfurt am Main, Germany}
\address{Institut für Theoretische Physik, Universität Frankfurt, Max-von-Laue-Straße 1, 60438 Frankfurt am Main, Germany}}
\maketitle

\begin{abstract}
  Dilepton production in heavy-ion collisions at various energies is
  studied using coarse-grained transport simulations. Microscopic output
  from the Ultra-relativistic Quantum Molecular Dynamics (UrQMD) model
  is hereby put on a grid of space-time cells which allows to extract
  the local temperature and chemical potential in each cell via an
  equation of state. The dilepton emission is then calculated applying
  in-medium spectral functions from hadronic many-body theory and
  partonic production rates based on lattice calculations. The
  comparison of the resulting spectra with experimental data shows that
  the dilepton excess beyond the decay contributions from a hadronic
  cocktail reflects the trajectory of the fireball in the
  $T-\mu_{\mathrm{B}}$ plane of the QCD phase diagram.
\end{abstract}

\eqsec
 
\section{Introduction}

Lepton pairs represent excellent probes for the properties of hot and
dense matter created in heavy-ion collisions. Their cross section for
interactions is extremely small so that they leave the produced fireball
unscathed. In addition, dileptons directly couple to the vector current
and therefore give access to the corresponding iso-scalar and iso-vector
spectral functions, which allows one to learn about (i) the properties
of hadrons in the medium, (ii) the predicted deconfinement and (iii)
chiral symmetry restoration \cite{Rapp:1999ej}.

However, the interpretation of experimental dilepton spectra is
challenging. Our understanding of the strong interaction, the properties
of matter and---in consequence---the processes driving the dynamics in a
heavy-ion collision is still limited. Because a full solution of QCD
from first principles is not possible yet, theory has to rely on the use
of models. Basically there exist two different approaches for the
description of heavy-ion collisions, and both have some advantages and
disadvantages: While in macroscopic models such as fireball
parameterizations or hydrodynamics the implementation of medium effects
is straightforward, they require a short mean free path of the hadrons;
in consequence they are only applicable at sufficiently high collision
energies and only for the hot and dense stage of the reaction. On the
other hand, microscopic transport approaches account for the individual
hadron-hadron interactions at all stages of the collision, but effects of finite 
temperature and density as, e.g., spectral modifications or phase 
transitions are difficult to implement.

One option to further improve the theoretical description is to
implement effective solutions for the full non-equilibrium quantum
transport problem \cite{KadanoffBaym1962}. However this is an extremely
difficult task which requires full self-consistency. The second path is
to connect the microscopic and macroscopic descriptions and combine the
advantages of both pictures. Following the latter idea, the goal of the
present work is to fully determine the macroscopic evolution from an
underlying microscopic picture.

\section{Coarse-graining approach}

Based on the previous work by Huovinen and collaborators
\cite{Huovinen:2002im}, the coarse-graining approach strongly simplifies
the description obtained from the microscopic dynamics by reducing the
information to a few thermodynamic quantities. As only a summarizing
sketch of the approach can be given here, we refer for further details
to Ref.~\cite{Endres:2014zua}.

\begin{figure}
\begin{center}
\includegraphics[width=0.68\columnwidth]{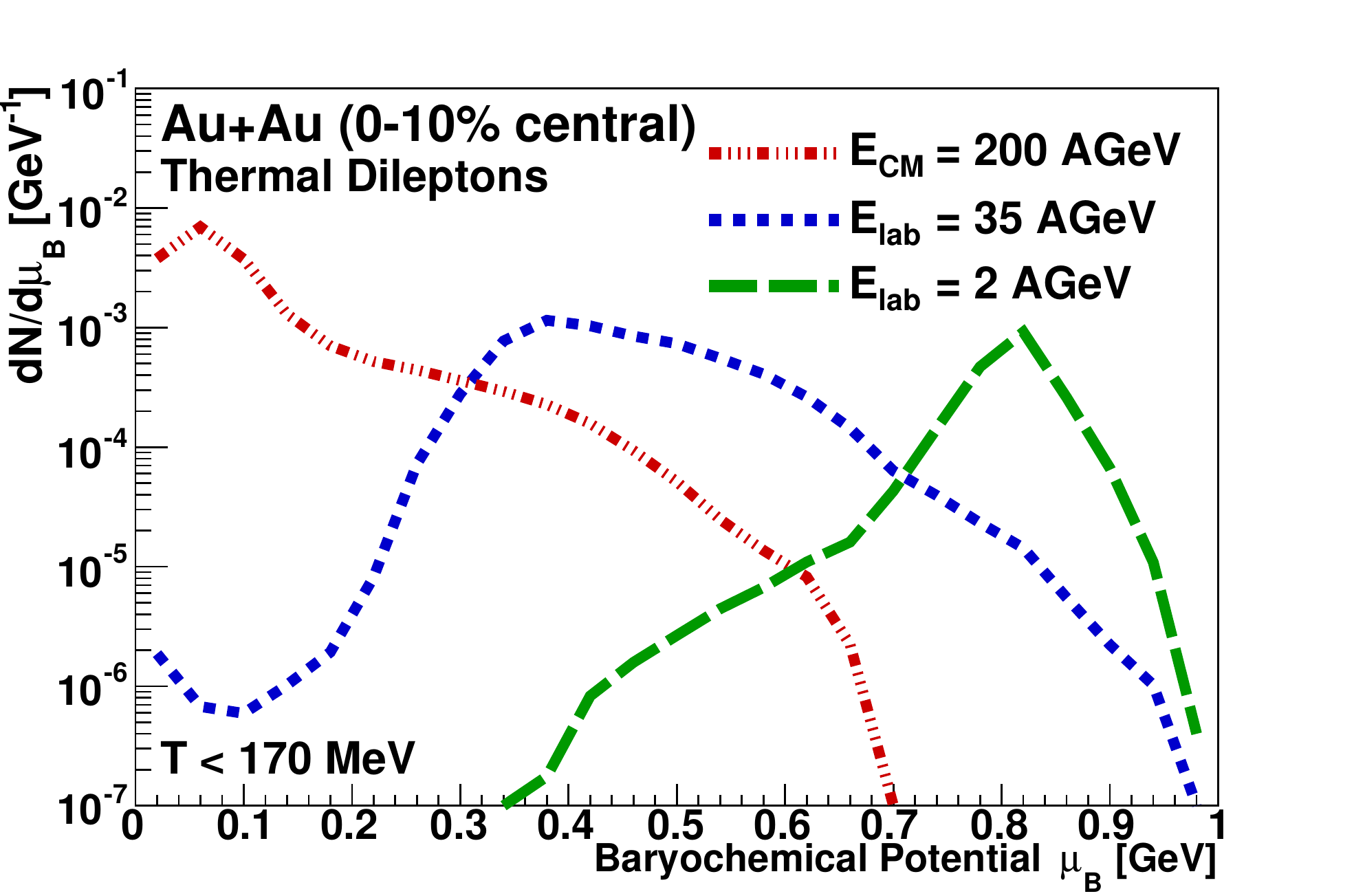}
\caption{Thermal dilepton emission in dependence on
  $\mu_{\mathrm{B}}$. The results are shown for central Au+Au collisions
  at the FAIR energies $E_{\mathrm{lab}}=2$ and $35\,A$GeV as well as
  for the top RHIC energy $\sqrt{s}=200\,A$GeV. Only the hadronic
  emission, i.e., the yield for $T < 170$\,MeV, is considered.}
\label{yieldmub}
\end{center}
\end{figure}

In general, the model can be subdivided into three steps as follows:
\begin{enumerate}
\item A large ensemble of events is calculated with the UrQMD transport
  model \cite{Bass:1998sa}, which describes the positions and momenta of all hadrons in the
  system for each time-step. A sufficient number of events is necessary
  to obtain a smooth distribution function $f(x,p,t)$. The output is
  then put on a space-time grid of small cells, which allows to
  determine the local baryon four-flow $j^{\mu}_{\mathrm{B}}$ and the
  energy-momentum tensor $T^{\mu\nu}$. Applying Eckart's requirement of
  vanishing baryon flow, one can define the rest-frame for each cell and
  determine the local energy and baryon density, $\varepsilon$ and
  $\rho_{\mathrm{B}}$.

\item To obtain the temperature $T$ and chemical potential $\mu_{\mathrm{B}}$ in each cell an equation of
  state (EoS) is required, which relates the thermodynamic quantities to
  the local densities $\varepsilon$ and $\rho_{\mathrm{B}}$. We here use
  (i) a hadron-gas EoS \cite{Zschiesche:2002zr} for $T \leq 170\,$MeV and 
  (ii) an EoS fitted to lattice calculations \cite{He:2011zx} for higher 
  temperatures. While the hadron-gas EoS provides good consistency with the 
  underlying microscopic dynamics because it includes the same degrees of freedom 
  as the UrQMD model, the lattice EoS is necessary for a correct description of the
  deconfined phase above the critical temperature. To account for
  deviations from chemical equilibrium, also the local pion chemical potential
  is extracted.
\item With given $T$ and $\mu_{\mathrm{B}}$, one can calculate the
  thermal dilepton emission for each cell. In a system in equilibrium
  the production rate is determined by the imaginary part of the
  electromagnetic current-current correlator $\Pi^{(\mathrm{ret})}$,
  which is (in the hadronic phase) related to the spectral distributions
  of the light vector mesons up to $M=1$\,GeV/$c^{2}$, according to the
  current-field identity. To include medium effects we use
  state-of-the-art spectral functions for the $\rho$ and $\omega$ meson
  from hadronic many-body theory \cite{Rapp:1999us}. At higher masses
  $\Pi^{(\mathrm{ret})}$ is dominated by a continuum of multi-meson
  states, for which the rates are obtained using a chiral-reduction
  approach \cite{vanHees:2007th}. In the QGP phase the main contribution
  to the dilepton production is given by quark-antiquark
  annihilation. Here we apply thermal emission rates obtained from
  lattice calculations \cite{Ding:2010ga}.
\end{enumerate}

\section{Summary of results}
\begin{figure}
\begin{center}
\includegraphics[width=0.68\columnwidth]{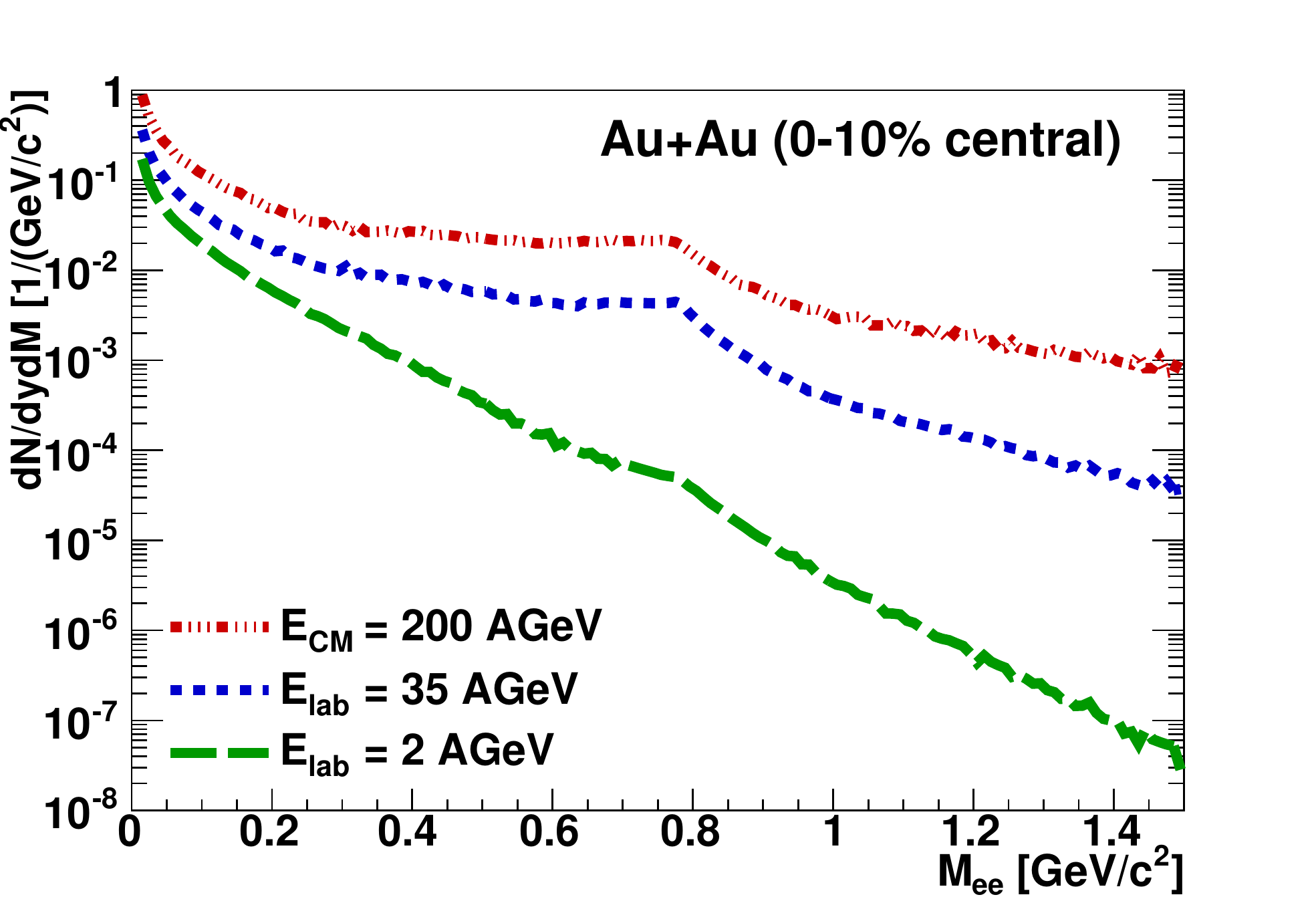}
\caption{Thermal dilepton invariant mass spectra $\dd N/(\dd y\, \dd M)$
  for central Au+Au collisions and the same energies as considered considered in
  Fig.\,\ref{yieldmub}. The results show the yield at mid-rapidity,
  $\vert y_{\mathrm{ee}} \vert < 0.5$.}
\label{theryieldcomp}
\end{center}
\end{figure}

Dilepton production from SIS\,18 up to LHC energies has been studied in
detail with the coarse-graining approach; the corresponding results
(compared to experimental data) can be found in
Ref.~\cite{Endres:2014zua}. They show that at all energies the
measured dilepton excess over the hadronic decay cocktail can be
interpreted as an effect of medium-modified spectral distributions, reflecting the emission conditions. In these
proceedings, we focus on the connection between the trajectory of the
fireball in the QCD phase diagram and the resulting dilepton spectra.

In Fig.\,\ref{yieldmub} the thermal hadronic dilepton yield for central
Au+Au collisions in relation to the value of $\mu_{\mathrm{B}}$ at the
source (i.e., in the emitting cell) is shown for different collision
energies from FAIR to RHIC. Within this energy range one observes a
strong variation of the baryochemical potential: While for
$E_{\mathrm{lab}}=2\,A$GeV one finds a clear peak of the thermal
dilepton emission for $\mu_{\mathrm{B}}=0.7-0.9$\,GeV, most lepton pairs
at RHIC stem from cells with (nearly) vanishing baryochemical
potential. Consequently, strong baryonic effects on the spectral shapes
of the vector mesons can be expected for low FAIR energies, while their
influence gradually decreases when going to higher collision
energies. At the same time, the temperature increases from a maximum
value of $T \approx 100$\,MeV at the lower FAIR energies up to
$400-500$\,MeV for $\sqrt{s}=200\,A$GeV at RHIC (not shown here).

How do the thermodynamic properties of the fireball show up in the
dilepton spectra? The corresponding thermal invariant-mass
yields are presented in Fig.\,\ref{theryieldcomp}. For the lowest collision
energy $E_{\mathrm{lab}}=2\,A$GeV the influence of the large $\mu_{\mathrm{B}}$ is obvious; the
spectrum shows a significant enhancement at low masses resulting in a
Dalitz-like shape. This is caused by the strong coupling of the
$\rho$ meson to the baryonic $\Delta$ and $\mathrm{N}^{*}_{1520}$
resonances. For the top FAIR energy $E_{\mathrm{lab}}=35\,A$GeV the
spectrum is much flatter due to decreasing baryonic influence
and higher temperatures. Especially the increase of $T$ results in a larger
contribution at higher masses. This trend continues for the top RHIC energy. However, the shape of the spectrum shows only a small
qualitative change, although the fraction of partonic emission from the
QGP increases strongly between FAIR and RHIC energies. The present results indicate that the thermal emission rates from hadronic and
partonic matter are dual to a large extent, resulting in a conformable
spectral shape. Thus the identification of signals for a phase
transition in the spectra requires systematic and precise studies, which remains a challenge for
the future experiments at FAIR and the beam-energy scan program at RHIC.

\section*{Acknowledgments}
The authors acknowledge Ralf Rapp for providing the parameterization of
the spectral functions. This work was supported by the BMBF, HIC for
FAIR, and H-QM.  
\bibliographystyle{num-hvh} 
\bibliography{Bibliothek}
\end{document}